\documentclass[preprint,showpacs,preprintnumbers,amsmath,amssymb]{revtex4}
\usepackage{graphicx}
\usepackage{dcolumn}
\usepackage{bm}
\usepackage{psfrag}

\begin{document}
\title{Correlation functions in the Non Perturbative Renormalization Group and field expansion}

\author{Diego Guerra}
  \email{dguerra@fing.edu.uy}
  \affiliation{Instituto de F\'{\i}sica, Facultad de Ingenier\'{\i}a, Univ.~de la Rep\'ublica, J.H.y Reissig 565, 11000
  Montevideo, Uruguay}
  \author{Ram\'on M\'endez-Galain}
  \email{mendezg@fing.edu.uy}
  \affiliation{Instituto de F\'{\i}sica, Facultad de Ingenier\'{\i}a, Univ.~de la Rep\'ublica, J.H.y Reissig 565, 11000
  Montevideo, Uruguay}
\author{Nicol\'as Wschebor}
  \email{nicws@fing.edu.uy}
  \affiliation{Instituto de F\'{\i}sica, Facultad de Ingenier\'{\i}a, Univ.~de la Rep\'ublica, J.H.y Reissig 565, 11000
  Montevideo, Uruguay}

  \date{\today}

\begin{abstract}
The usual procedure of including a finite number of vertices in Non Perturbative Renormalization Group equations in order to obtain $n$-point correlation functions at finite momenta is analyzed.
This is done by exploiting a general method recently introduced which includes simultaneously all vertices 
although approximating their momentum dependence. The study is performed using the self-energy of the tridimensional scalar model at criticality.
At least in this example, low order truncations miss quantities as the critical exponent $\eta$ by as much as 60\%. However, if one goes to high order truncations the procedure seems to converge rapidly.
\end{abstract}

\maketitle


\def\bfphi{\mbox{\boldmath$\phi$}}
\def\bfvarphi{\mbox{\boldmath$\varphi$}}
\def\bfgamma{\mbox{\boldmath$\gamma$}}
\def\bfalpha{\mbox{\boldmath$\alpha$}}
\def\bftau{\mbox{\boldmath$\tau$}}
\def\bfnabla{\mbox{\boldmath$\nabla$}}
\def\bfsigma{\mbox{\boldmath$\sigma$}}
\def\bfpi{\mbox{\boldmath$\pi$}}

\newcommand \beq{\begin{eqnarray}}
\newcommand \eeq{\end{eqnarray}}

\newcommand \ga{\raisebox{-.5ex}{$\stackrel{>}{\sim}$}}
\newcommand \la{\raisebox{-.5ex}{$\stackrel{<}{\sim}$}}

\def\psib{\psi}
\def\phib{\phi}
\def\r{{\rm r}}
\def\d{{\rm d}}

\def \e {\mbox{e}}

\input epsf


\def\square{\hbox{{$\sqcup$}\llap{$\sqcap$}}}
\def\grad{\nabla}
\def\del{\partial}

\def\frac#1#2{{#1 \over #2}}
\def\smallfrac#1#2{{\scriptstyle {#1 \over #2}}}
\def\half{\ifinner {\scriptstyle {1 \over 2}}
   \else {1 \over 2} \fi}


\def\bra#1{\langle#1\vert}
\def\ket#1{\vert#1\rangle}


\def\simge{\mathrel{%
   \rlap{\raise 0.511ex \hbox{$>$}}{\lower 0.511ex \hbox{$\sim$}}}}
\def\simle{\mathrel{
   \rlap{\raise 0.511ex \hbox{$<$}}{\lower 0.511ex \hbox{$\sim$}}}}


\def\buildchar#1#2#3{{\null\!
   \mathop#1\limits^{#2}_{#3}
   \!\null}}
\def\overcirc#1{\buildchar{#1}{\circ}{}}


\def\slashchar#1{\setbox0=\hbox{$#1$}
   \dimen0=\wd0
   \setbox1=\hbox{/} \dimen1=\wd1
   \ifdim\dimen0>\dimen1
      \rlap{\hbox to \dimen0{\hfil/\hfil}}
      #1
   \else
      \rlap{\hbox to \dimen1{\hfil$#1$\hfil}}
      /
   \fi}


\def\real{\mathop{\rm Re}\nolimits}     
\def\imag{\mathop{\rm Im}\nolimits}     

\def\tr{\mathop{\rm tr}\nolimits}       
\def\Tr{\mathop{\rm Tr}\nolimits}       
\def\Det{\mathop{\rm Det}\nolimits}     

\def\mod{\mathop{\rm mod}\nolimits}     
\def\wrt{\mathop{\rm wrt}\nolimits}     


\def\TeV{{\rm TeV}}                     
\def\GeV{{\rm GeV}}                     
\def\MeV{{\rm MeV}}                     
\def\KeV{{\rm KeV}}                     
\def\eV{{\rm eV}}                       

\def\mb{{\rm mb}}                       
\def\mub{\hbox{$\mu$b}}                 
\def\nb{{\rm nb}}                       
\def\pb{{\rm pb}}                       

%
%

\def\picture #1 by #2 (#3){
  \vbox to #2{
    \hrule width #1 height 0pt depth 0pt
    \vfill
    \special{picture #3} 
    }
  }

\def\scaledpicture #1 by #2 (#3 scaled #4){{
  \dimen0=#1 \dimen1=#2
  \divide\dimen0 by 1000 \multiply\dimen0 by #4
  \divide\dimen1 by 1000 \multiply\dimen1 by #4
  \picture \dimen0 by \dimen1 (#3 scaled #4)}
  }

\def\centerpicture #1 by #2 (#3 scaled #4){
   \dimen0=#1 \dimen1=#2
    \divide\dimen0 by 1000 \multiply\dimen0 by #4
    \divide\dimen1 by 1000 \multiply\dimen1 by #4
         \noindent
         \vbox{
            \hspace*{\fill}
            \picture \dimen0 by \dimen1 (#3 scaled #4)
            \hspace*{\fill}
            \vfill}}


\def\figfermass{\centerpicture 122.4mm by 32.46mm
 (fermass scaled 750)}

%

\section{Introduction}

In nearly all fields in physics, there are systems having a large number of strongly correlated constituents.
These cannot be 
treated with usual perturbative methods. Phase transitions and critical phenomena, disordered systems, strongly 
correlated electrons, quantum chromodynamics at large distances, are just a few examples which demand a general and 
efficient method to treat non-perturbative situations. In problems as those just quoted, the calculation of 
correlation functions of the configuration variables is, in general, a very complicated task. 

The non perturbative renormalization group
(NPRG) \cite{Wilson73,Polchinski83,Wetterich93,Ellwanger93,Morris94} has proven to be a powerful tool to achieve this goal. It presents itself as an infinite hierarchy of flow equations relating sequentially the various $n$-point functions. 
It has been successfully applied in many different problems, either in condensed matter, particle or nuclear
physics (for  reviews, see e.g.
\cite{Bagnuls:2000ae,Berges02,Canet04}; a pedagogical introduction can be found in \cite{delamotte:2007}). 
In most of these problems however, one is interested in observables 
dominated by long wavelength modes. In these cases, it is then possible to approximately close the 
infinite hierarchy of NPRG equations performing an expansion in 
the number of derivatives of the field. This approximation scheme is known as the derivative expansion (DE) \cite{Golner86}. The price to pay is that the $n$-point functions can be calculated only at small external momenta, i.e. smaller than the smallest mass on the problem (vanishing momenta in the case of critical phenomena).

In many other physical problems however, this is
not enough: the full knowledge of the momentum dependence of 
correlation functions is needed in order to calculate quantities of
physical interest (e.g. to get the spectrum of excitations, the
shape of a Fermi surface, the scattering matrix, etc.). There have been many attempts to solve the
infinite system of flow equations at finite momenta; most of them are based on various forms of 
an early proposal by Weinberg \cite{weinberg73}. Although some of these attempts 
\cite{truncation,Ellwanger94,Ellwanger94a,Blaizot:2004qa} introduce sophisticated 
ansatz for the unknown correlation functions appearing in a given flow equation, 
most efforts simply ignore high order vertices.
In all these works, only low order vertices are taken into account: usual calculations do not even
include the complete flow of the $3$- and $4$-point functions. Moreover, it is not possible a priori to gauge the quality of such approximations schemes.

Recently, an alternative general method to get $n$-point functions at any finite momenta within the 
NPRG has been proposed 
\cite{BMW}. It has many similarities with DE. First, it is an approximation scheme that can be 
systematically improved. Second, the scheme yields a closed set of flow equations including simultaneously an infinite number of vertices; one thus goes far beyond schemes including a small number of vertices, as those quoted in the previous paragraph. Moreover, it has been proven \cite{BMW} that in their corresponding limits, 
both perturvative and DE results are recovered; this remains valid at each order of the 
respective expansion. Finally, in the large-$N$ limit 
of $O(N)$ models, the leading-order (LO) of the approximation scheme becomes exact for all $n$-point functions. 
(The expression ``leading order" 
means the first step in the approximation scheme; it does not refer 
to an expansion in a small parameter which usually does not exist in these kind of problems.). 

In \cite{BMW-num}, the method has been applied, in its leading order, to the calculation of the self-energy 
of the scalar model, at criticality. That is, we have fine-tuned the bare mass of the model in order for the correlation length to be infinite, and then we have studied the full range of momenta, from the high momenta Gaussian regime to the low momenta scaling one.
At this order of the approximation scheme the self-energy is expected to include all one loop contributions and to achieve DE at next-to-leading order (NLO) precision, in the corresponding limit \cite{BMW}. The numerical solution found in \cite{BMW-num} verifies these properties. Moreover, the function has the expected physical properties in all momenta regime. First, it presents the correct scaling behavior in the infrared limit. The model reproduces critical exponent $\eta$ with a level of precision comparable to the DE at NLO. Moreover, contrarily to DE, the anomalous power-law behavior can be read directly from the momentum dependence of the $2$-point function. Second, it shows the expected logarithmic 
shape of the perturbative regime even though the coefficient in front of the logarithm, which is a $2$-loops quantity, is only reproduced with an error of $8$\%. In order to check the quality of the solution in the intermediate momentum region, a quantity sensitive to this crossover sector has been calculated: one gets a result almost within the error bars of both Monte-Carlo and resumed $7$-loops perturbative calculations. Please observe that this quantity is extremely difficult to calculate: even these sophisticated methods give an error of the order of $10$\%. 

Another interesting similarity between DE and the method presented in \cite{BMW} 
is that, as a price to pay in order to close the equations including an infinite number of vertices, 
one has to study the problem in an 
external constant field. Accordingly, one ends
up with partial differential equations which may be difficult to solve. A useful approximation scheme, 
widely used in DE calculations, is to perform, on top of the expansion in  derivatives of the 
field, an extra expansion in powers of the field (see e.g.~\cite{Berges02}), in the spirit of Weinberg proposal. 
During the last 10 years, this strategy has been widely used  
\cite{expansion,Canet02,litimreg,Tetradis94,morris,Canet03,aoki,Litim-exp}; in many studied situations 
this expansion seems to converge (generally oscillating) \cite{Tetradis94,Canet02,litimreg,Canet03,aoki,Litim-exp}, while in many others it does not \cite{Tetradis92,morris}. In $d=2$, the field expansion has been explored with no indication
of convergence for critical exponents, even going to high orders \cite{private}.

In this work we shall explore this procedure of expansion in powers of the field, in the framework of 
the calculation scheme presented in \cite{BMW}. 
More precisely, we shall make a field expansion on top of the already approximated 2-point function flow equation
solved in \cite{BMW-num}. Then we shall compare results with and without field expansion.
In doing so, we have two goals. First, we shall study 
the apparent convergence of this procedure. This comparison is essential if one hopes to apply the scheme 
described in \cite{BMW} to situations more complicated than that considered in \cite{BMW-num}. For example, 
within DE scheme, when trying to go to higher orders or when considering more involved models, the expansion in powers of the field  on top of the corresponding approximate flow equations is sometimes
the only practical strategy to solve them \cite{Canet04,difficult-exp,Canet03}. The second and more important goal is the following: as we shall see in section \ref{expansion}, truncation in powers of the field is equivalent to ignoring high order vertices in the flow equations. Thus, the comparison presented here can help to estimate the quality of the calculations made so far to get $n$-point functions at finite momenta neglecting high order vertices.

The article is organized as follows. In the next section we describe the basics ingredients of both the NPRG 
and the  approximation scheme introduced in \cite{BMW}. We also present the results obtained in \cite{BMW-num}, when this scheme is used to find the 2-point function of the scalar model. In 
section \ref{expansion}, we apply the expansion in the field at various orders and compare these results 
with those found in \cite{BMW-num}. Finally, we present the conclusions of the study.

\section{General considerations}

\label{general}

Let us consider a scalar field theory with the classical action
\begin{equation}\label{eactON}
S = \int {\rm d}^{d}x\,\left\lbrace{ \frac{1}{2}}   \left(\del_\mu
\varphi(x)\right)^2  + \frac{r}{2} \, \varphi^2(x) + \frac{u}{4!}
\,\varphi^4(x) \right\rbrace \,.
\end{equation}
Here, $r$ and $u$ are the microscopic mass and coupling, respectively.

The NPRG builds a family of effective actions, 
$\Gamma_\kappa[\phi]$ (where $\phi(x)=\langle \varphi(x) \rangle_J$ is the expectation value of the 
field in presence of an external source $J(x)$),
in which  the magnitude of long wavelength fluctuations are
controlled by an infrared regulator  depending on a continuous
parameter $\kappa$. One can write for $\Gamma_\kappa[\phi]$ an exact
flow equation \cite{Tetradis94,Ellwanger94a,Morris94,Morris94c}:
\begin{equation}
\label{NPRGeq}
\partial_\kappa \Gamma_\kappa[\phi]=\frac{1}{2} \int \frac{d^dq}{(2\pi)^d}
\partial_\kappa R_\kappa(q^2)
\left[\Gamma_\kappa^{(2)}+R_\kappa\right]^{-1}_{q,-q},
\end{equation}
where $\Gamma_\kappa^{(2)}$ is the second functional derivative of
$\Gamma_\kappa$ with respect to $\phi(x)$,
and $R_\kappa$ denotes a family of ``cut-off functions''
depending on $\kappa$: 
$R_\kappa(q)$ behaves like $\kappa^2$ when  $q\ll \kappa$ and it vanishes 
rapidly when $q\gg \kappa$
\cite{litimreg,Canet02}. The effective action
$\Gamma_\kappa[\phi]$ interpolates between the classical action
obtained for $\kappa=\Lambda$ (where $\Lambda^{-1}$ is the microscopic length scale), and the full effective action obtained when  $\kappa \to 0$, i.e., when all fluctuations
are taken into account (see e.g. \cite{Berges02}).

By differentiating eq.~(\ref{NPRGeq}) with respect to $\phi(x)$, and then
letting the field be constant, one gets the flow equation for the
$n$-point function $\Gamma_\kappa^{(n)}$ in a constant background field
$\phi$. 
  For example, for the 2-point function one gets:
\begin{eqnarray}
\label{gamma2champnonnul}
\partial_\kappa\Gamma_{\kappa}^{(2)}(p;\phi)&=&\int
\frac{d^dq}{(2\pi)^d}\partial_\kappa R_k(q)\Big\{G_{\kappa}(q;\phi)
\Gamma_{\kappa}^{(3)}(p,q,-p-q;\phi) 
\nonumber \\
&&\times G_{\kappa}(q+p;\phi)\Gamma_{\kappa}^{(3)}(-p,p+q,-q;\phi)
G_{\kappa}(q;\phi) 
\nonumber  \\
&&-\frac{1}{2}G_{\kappa}(q;\phi)\Gamma_{\kappa}^{(4)}
(p,-p,q,-q;\phi)G_{\kappa}(q;\phi)\Big\} ,
\end{eqnarray}
where
\begin{equation}\label{G-gamma2}
G^{-1}_{\kappa} (q;\phi) \equiv \Gamma^{(2)}_{\kappa} (q,-q;\phi)
+ R_\kappa(q^2),
\end{equation}
and we used the definition 
\begin{equation}\label{gamman}
(2\pi)^d \;\delta^{(d)}\left(\sum_i 
p_i\right)\;\Gamma_\kappa^{(n)}(p_1,\dots,p_n;\phi)= \int 
d^dx_1\dots\int d^dx_{n}
e^{i\sum_{j=1}^n p_jx_j}\left. \frac{\delta^n\Gamma_\kappa}{\delta\phi(x_1)
\dots \delta\phi(x_n)}\right|_{\phi(x)\equiv \phi}.
\end{equation}

The flow equation for a given $n$-point function involves 
the $n+1$ and $n+2$ point functions (see,
e.g., eq.~(\ref{gamma2champnonnul})), so that
the flow equations for all correlation functions constitute an
infinite hierarchy of coupled equations.

In \cite{BMW}, a general method to solve this infinite hierarchy was proposed.
It exploits the smoothness of the regularized $n$-point functions, 
and the fact that the loop momentum $q$ in the right
hand side of the flow equations (such as eq.~(\ref{NPRGeq}) or
eq.~(\ref{gamma2champnonnul})) is limited to  $q\simle \kappa$ due to
the presence of $\partial_\kappa R_\kappa(q)$. The leading order 
of the method presented in \cite{BMW} thus consists in setting
\beq\label{BMWLO}
\Gamma^{(n)}_{\kappa}(p_1,p_2,...,p_{n-1}+q,p_n-q)\sim
\Gamma^{(n)}_{\kappa}(p_1,p_2,...,p_{n-1},p_n)
\eeq
in the r.h.s. of the flow equations. After making this approximation, some momenta in
some of the $n$-point functions vanish, and their expressions can then be obtained as derivatives of $m$-point
functions ($m<n$) with respect to a constant background field.

Specifically, in the flow equation for the 2-point function, eq.~(\ref{gamma2champnonnul}), after setting $q=0$ in the vertices of the r.h.s., the 3- and 4-point functions will contain one
and two vanishing momenta, respectively. These  can be related
to the following derivatives of the 2-point function:
\begin{equation}\label{derivs}
\Gamma_{\kappa}^{(3)}(p,-p,0;\phi)=\frac{\partial
\Gamma_{\kappa}^{(2)} (p,-p;\phi)} {\partial \phi} , \hskip 1 cm
\Gamma_{\kappa}^{(4)}(p,-p,0,0;\phi)=\frac{\partial^2
\Gamma_{\kappa}^{(2)} (p,-p;\phi)} {\partial \phi^2}.
\end{equation}
One then gets a closed equation for
$\Gamma_{\kappa}^{(2)}(p;\phi)$:
\begin{equation}
  \label{2pointclosed}
\kappa \partial_\kappa\Gamma_\kappa^{(2)}(p^2;\phi)=
J_d^{(3)}(p,\kappa;\phi) \; \left( \frac{\partial
\Gamma_\kappa^{(2)}(p,-p;\phi)} {\partial \phi} \right)^2
   -\frac{1}{2} I_d^{(2)}(\kappa;\phi) \; \frac{\partial^2
\Gamma_\kappa^{(2)}(p,-p;\phi)} {\partial \phi^2},
\end{equation}
where 
\beq\label{defJ} 
J_d^{(n)}(p;\kappa;\phi)\equiv
\int\frac{d^dq}{(2\pi)^d}\kappa \partial_\kappa R_\kappa(q^2)
G_\kappa(p+q;\phi)G^{(n-1)}_\kappa(q;\phi) , 
\eeq 
and
\beq\label{defI} 
I_d^{(n)}(\kappa;\phi)\equiv \int
\frac{d^dq}{(2\pi)^d}\kappa \partial_\kappa R_\kappa(q^2)
G^n_\kappa(q;\phi). 
\eeq  
In fact, in order to preserve the relation
\beq
\Gamma^{(2)}_{\kappa}(p=0;\phi) = \frac{\partial^2 V_\kappa}{\partial \phi^2},
\eeq
$V_\kappa(\phi)=\Gamma_\kappa[\phi(x)\equiv\phi]/\mathrm{Vol}$ being the effective potential, 
it is better to make the approximation (\ref{BMWLO}) (followed by (\ref{derivs})) in the flow equation for 
$\Sigma_\kappa(p;\phi)$
defined as
\begin{equation}\label{def-sigma}
 \Sigma_\kappa (p;\phi) = \Gamma^{(2)}_{\kappa} (p;\phi) - p^2 - \Gamma^{(2)}_{\kappa} (p=0;\phi).
\end{equation}
The $2$-point function is then obtained from $\Gamma^{(2)}(p;\phi)=\partial^2 V_\kappa(\phi)/\partial \phi^2 + p^2 + 
\Sigma_\kappa(p;\phi)$, which demands the simultaneous solution of the flow equations for $V_\kappa(\phi)$ and 
$\Sigma_\kappa (p;\phi)$.

As shown in \cite{BMW-num}, even if the complete solution of these equations is a priori complicated, a simple, and still accurate, way of solving them consists in assuming in the various 
integrals
\beq
G^{-1}_{\kappa} (q;\phi) \simeq Z_\kappa q^2 + \partial^2 V_\kappa(\phi)/\partial \phi^2 + R_\kappa(q^2),
\eeq
where $Z_\kappa \equiv Z_\kappa(\phi=0)$, with $Z_\kappa(\phi)\equiv 1+ \partial \Sigma_\kappa(p;\phi) / \partial 
p^2|_{p=0}$. This approximation is consistent with an improved version of the Local Potential Approximation 
(LPA, the first 
order of the DE), which includes explicitly a field renormalization factor $Z_\kappa$ \cite{Berges02}. 
Doing so, the ``$p=0$'' sector decouples from the $p\neq 0$ one. Here, by ``$p=0$'' we mean  the sector
describing vertices and derivative of vertices at zero momenta, i.e., flow equations for $V_\kappa$ and $Z_\kappa$. 
Moreover, it is useful to use the regulator \cite{litimreg}
\beq \label{reg-litim}
R_\kappa(q^2) =Z_\kappa (\kappa^2-q^2) \; \Theta(\kappa^2-q^2),
\eeq
which allows the functions $J_d^{(n)}(p;\kappa;\phi)$ and $I_d^{(n)}(\kappa;\phi)$ to be calculated 
analytically. The corresponding expressions can be found in \cite{BMW-num}.
In fact, all quantities are functions of $\rho\equiv \phi^2/2$.
The problem is then reduced to the solution of the three flow equations for $V_\kappa(\rho)$  and 
$Z_\kappa(\rho)$, for the $p = 0$ sector, and for $\Sigma_\kappa(p;\rho)$, in the $p\neq 0$ one. 
As only the ``effective mass'' 
\beq
m^2_\kappa(\rho) \equiv \frac{\partial^2 V_\kappa (\phi)}{\partial \phi^2} = \frac{\partial V_\kappa (\rho)}{\partial \rho} + 2\rho  \; \frac{\partial^2 V_\kappa (\rho)}{\partial \rho^2}
\eeq
(and its derivatives with respect to $\rho$) enters in the 
$p\neq 0$ sector, it is more convenient to work with the flow equation for $m^2_\kappa (\rho)$ instead of that 
for $V_\kappa(\rho)$ itself. The non-trivial fact is that by differentiating twice the flow equation for $V_\kappa(\rho)$ 
w.r.t $\phi$, one gets a closed equation for $m^2_\kappa (\rho)$.

In order to make explicit the fixed point in the $\kappa \to 0$ limit, it is necessary to work with 
dimensionless variables:
\beq
\mu_\kappa(\tilde \rho) \equiv Z_\kappa^{-1} \; \kappa^{-2} \; m^2_\kappa(\rho)  
\hskip 0.3 cm ,\hskip 1 cm 
 \chi_\kappa(\tilde \rho) \equiv Z_\kappa^{-1} \; Z_\kappa(\rho)
\hskip 0.3 cm , \hskip 1 cm \tilde \rho \equiv K_d^{-1} \; Z_\kappa \; \kappa^{2-d} \; \rho \hskip 0.3 cm ,
\eeq
which, in the critical case, have a finite limit when $\kappa \to 0$.
Above, $K_d$ is a constant conveniently taken as 
$K_d^{-1}\equiv d\; 2^{d-1}\; \pi^{d/2} \; \Gamma(d/2)$ (e.g., $K_3=1/(6\pi^2)$). In the $p\neq 0$ sector, 
the dimensionful variable $p$ in the self-energy flow equation makes $\Sigma_\kappa(p;\tilde \rho)$ 
reach a finite value 
when $\kappa \to 0$. As discussed in \cite{BMW-num}, the inclusion of the 
flow equation for the renormalization factor $Z_\kappa(\tilde \rho)$ is essential in order to 
preserve the correct scaling behavior of $\Gamma^{(2)}(p;\tilde \rho)$ in the infrared limit. Doing so,
in the critical case, the function $\Gamma^{(2)}(p;\tilde \rho)/(Z_\kappa \kappa^2)$ has to reach a fixed point
expression depending on $\tilde \rho$ and $p/\kappa$, when $\kappa, p \ll u$ and $\tilde \rho \sim 1$.

Putting all together, in $d=3$, the three flow equations that have to be solved are
\beq
\kappa \partial_\kappa\mu_\kappa(\tilde{\rho})=-(2-\eta_\kappa)\mu_\kappa(\tilde{\rho})
+(1+\eta_\kappa)\tilde{\rho}\mu_\kappa'(\tilde{\rho})-\left(1-\frac{\eta_\kappa}{5}\right)
\left(\frac{\mu_\kappa'(\tilde{\rho})+2\tilde{\rho}\mu_\kappa''(\tilde{\rho})}
{(1+\mu_\kappa(\tilde{\rho}))^2}-\frac{4\tilde{\rho}\mu_\kappa'(\tilde{\rho})^2}
{(1+\mu_\kappa(\tilde{\rho}))^3}\right)  \nonumber \\
\label{eqmasa}
\eeq
and
\beq
&\kappa \partial_\kappa\chi_\kappa(\tilde \rho)= \eta_\kappa \chi_\kappa(\tilde \rho) + (1+\eta_\kappa) \tilde \rho \chi_\kappa'(\tilde \rho) 
- 2 \tilde \rho
\frac{\mu_\kappa'^2(\tilde \rho)}{(1+\mu_\kappa(\tilde \rho))^4} \nonumber \\
&\hskip 1.0 cm +\left( 1-\frac{\eta_\kappa}{5}\right) \left( 8\tilde \rho \chi'_\kappa(\tilde \rho) \frac{\mu'_\kappa(\tilde \rho)} {(1+\mu_\kappa(\tilde \rho))^3}
- \frac{\chi'_\kappa(\tilde \rho)+2\tilde\rho\chi''_\kappa(\tilde \rho)}
{(1+\mu_\kappa(\tilde \rho))^2}\right),
\label{eqchi}
\eeq
together with
\beq
\eta_\kappa=\frac{\chi_\kappa'(0)}{\chi_\kappa'(0)/5+(1+\mu_\kappa(0))^2}\hskip 0.2 cm , \hskip 1cm
\kappa \partial_\kappa Z_{\kappa}=-\eta_\kappa Z_{\kappa} \hskip 0.2 cm ,
\eeq
for the $p=0$ sector, and
\begin{eqnarray}
&\hspace{-2cm}\kappa \partial_\kappa\Sigma_\kappa(p,\tilde{\rho})=(1+\eta_\kappa)\tilde{\rho}\Sigma_\kappa'(p,\tilde{\rho})
+\frac{2\tilde{\rho}\mu_\kappa'^2(\tilde{\rho})\kappa^2Z_\kappa}{(1+\mu_\kappa(\tilde{\rho}))^2}
\left(f_\kappa(\tilde{p},\tilde{\rho})-\frac{2(1-\eta_\kappa/5)}{(1+\mu_\kappa(\tilde{\rho}))^2} \right) 
\nonumber \\
&+ \frac{2\tilde{\rho} f_\kappa(\tilde{p},\tilde{\rho})}{(1+\mu_\kappa(\tilde{\rho}))^2} \; 
\left(2\mu_\kappa'(\tilde{\rho})\Sigma_\kappa'(p,\tilde{\rho})+\frac{\Sigma_\kappa'^2(p,\tilde{\rho})}{\kappa^2
Z_\kappa}\right)-\frac{(1-\eta_\kappa/5)}{(1+\mu_\kappa(\tilde{\rho}))^2}\left(\Sigma_\kappa'(p,\tilde{\rho})
+2\tilde{\rho}\Sigma_\kappa''(p,\tilde{\rho})\right)
\label{eqsigma}
\end{eqnarray}
for the $p\neq 0$ one. In these equations, the prime means $\partial_{\tilde \rho}$ and we used the 
explicit expression for $I_3^{(n)}=2 K_3 {\kappa^{5-2n}}{Z_\kappa^{1-n}}(1-{\eta_\kappa}/5)
/{(1+\mu_\kappa(\tilde \rho))^n} $. In eq.~(\ref{eqsigma}), we introduced 
the dimensionless expression $f_\kappa$
defined as $J_3^{(3)}(p;\kappa;\rho) \equiv K_3 \kappa^{-1} Z^{-2}_\kappa / (1+\mu_\kappa (\tilde \rho))^2 
\times f_\kappa(\tilde p;\tilde\rho)$, with $\tilde p \equiv p/\kappa$.

In \cite{BMW-num}, this strategy is used to get the $2$-point function of the scalar model at criticality  and 
zero external field (i.e., $\Sigma(p=0,\rho=0)=0$), in $d=3$. As recalled above, the function thus obtained 
has the correct shape, either in the scaling, perturbative and intermediate momenta regimes.

\section{Expansion in powers of the field}
\label{expansion}

\begin{figure}[t]
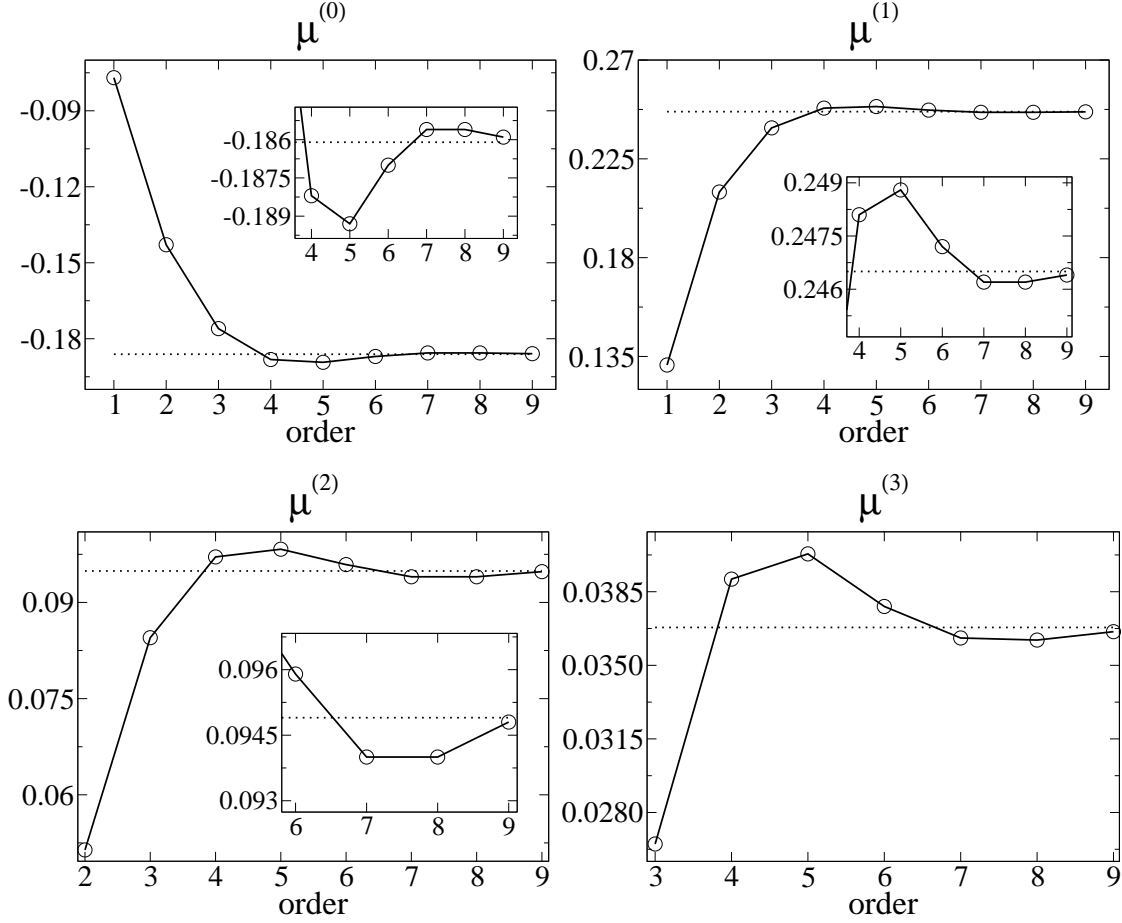

\begin{center}
\includegraphics*[scale=0.29]
{figmu0b.eps}
\includegraphics*[scale=0.29]
{figmu1b.eps}
\\[10pt]
\includegraphics*[scale=0.29]
{figmu2b.eps}
\includegraphics*[scale=0.29]
{figmu3b.eps}
\end{center}
\caption{ \label{exp-LPA} First four dimensionless fixed point couplings at zero momenta and zero external field: results 
obtained by truncating the flow equation, as a function of the order; the corresponding value for the complete
equation is represented by the dotted-line.}
\end{figure}

In this section, we shall compare the solution obtained in \cite{BMW-num} using the procedure described above, 
with the solution of the same three flow equations expanded in powers of $\tilde \rho$ and truncating up to a given order. 
Before doing so, let us first consider only the flow equation for the potential or, equivalently, that
for the effective mass, i.e., eq.~(\ref{eqmasa}), with $Z_\kappa\equiv 1$ ($\eta_\kappa \equiv 0$). This corresponds to the pure LPA sector and it is thus independent of the  
scheme presented in \cite{BMW}. In $d=3$, its expansion in powers of the field has been widely studied 
during the last ten years, using various regulators \cite{morris,aoki}.
Recently, another interesting truncation scheme has also been considered in \cite{Boisseau06}
showing much better convergence properties. However, here we shall consider the simpler expansion in powers of the fields; as shall be seen bellow this is the field expansion that can be compared to usual truncation in the number of vertices.
It has been shown \cite{Litim-exp} that, using the regulator we 
consider here (see eq.~(\ref{reg-litim})), this expansion seems to converge. 
This result follows when expanding both around finite and zero external field, although faster in the first case.
In \cite{Litim-exp} the convergence in this situation has been discussed studying the critical exponent $\nu$.
In order to strengthen this conclusion, as a first step in our study, we have analyzed the effect 
of the expansion on the function $\mu_\kappa(\tilde \rho)$:
\beq
\mu_\kappa(\tilde \rho) = \sum_{n=0}^\infty \frac{1}{n!}\; \mu_\kappa^{(n)} \; {\tilde \rho}^n.
\label{expmu}
\eeq 
More precisely, we shall gauge the impact of truncating this sum on the fixed point values of the coefficients $\mu_{\kappa}^{(n)}$, which are proportional to vertices at zero momenta and zero external field. This study is motivated by the fact that these $\mu_{\kappa}^{(n)}$ shall appear in the $\Sigma_\kappa(p;\tilde \rho)$ 
flow equation, eq.~(\ref{eqsigma}),
when the later shall be expanded around $\tilde \rho=0$. Results are shown in Figure \ref{exp-LPA}. 
The four plots present the fixed point value for the first 4 couplings, 
$\mu_{\kappa=0}^{(n)}$, $n=0, \cdots , 3$.
For each coupling, we present the result which follows by solving the complete LPA equation, eq.~(\ref{eqmasa}), 
together with the result 
obtained with the equation expanded in powers of $\tilde \rho$. For example, when going only up to the first 
order (i.e., neglecting all $\mu_\kappa^{(n)}$ with $n\geq 2$), the corresponding equations for $\mu_\kappa^{(0)}$ and $\mu_\kappa^{(1)}$, are:
\beq
\kappa \partial_\kappa\mu_\kappa^{(0)}=(\eta_\kappa-2)\mu_\kappa^{(0)}
-\frac{(1-\eta_\kappa/5)\mu_\kappa^{(1)}}{(1+\mu_\kappa^{(0)})^2}
\label{expmu0}
\eeq
and
\beq	
\kappa\partial_\kappa\mu_\kappa^{(1)}=(2\eta_\kappa-1)\mu_\kappa^{(1)}+\frac{6(1-\eta_\kappa/5)
{(\mu^{(1)}_\kappa)}^2}{(1+\mu_\kappa^{(0)})^3}.
\label{expmu1}
\eeq
which have to be solved simultaneously. (In fact, if solving just the LPA, $\eta_\kappa=0$; nevertheless, we have kept $\eta_\kappa$ in eqs.~(\ref{expmu0})-(\ref{expmu1}) for a later use of these equations).
When going to the second order, eq.~(\ref{expmu1}) acquires 
a new term and 
a new flow equation, that for $\mu_\kappa^{(2)}$, appears; and so on.
According to Figure \ref{exp-LPA}, an apparent convergence 
shows up. In all cases one observes that: 1) there seems to be an oscillating convergence, 2) the value of
$\mu^{(i)}$ is found with about 1\% error truncating at order $i+3$.

Let us now turn to the study of the flow equation for the $2$-point function coming from the scheme 
proposed in \cite{BMW}. As the effective potential (or the effective mass), 
$ \Gamma^{(2)}_\kappa(p;\rho)$ can also be expanded in powers of the external field:
\beq\label{exp-2p}
\Gamma^{(2)}_\kappa(p;\rho) = \sum_{n=0}^\infty \frac{2^n}{(2n)!} \;
\Gamma^{(2n+2)}_\kappa(p,-p,0,0,\cdots,0;\rho)|_{\rho=0}\; \; \rho^n ,
\eeq
because
\beq\label{der2p}
\Gamma^{(m+2)}_\kappa(p,-p,0,0,\cdots,0;\rho) = \frac{\partial^{m}\Gamma_\kappa^{(2)}(p;\phi)}{\partial \phi^{m}}
\eeq
and we used that, at zero field, all odd vertex functions vanish. Equation (\ref{exp-2p}) makes clear the 
point stated above: once approximation (\ref{BMWLO}) is performed, truncating the expansion in powers of the 
external field is equivalent to neglecting high 
order vertices. Moreover, eqs.~(\ref{exp-2p}) and (\ref{der2p}) show that the procedure proposed in  
\cite{BMW} indeed includes all vertices, although approximately.

\begin{figure}[t]
\begin{center}
\includegraphics*[scale=0.4,angle=-90]{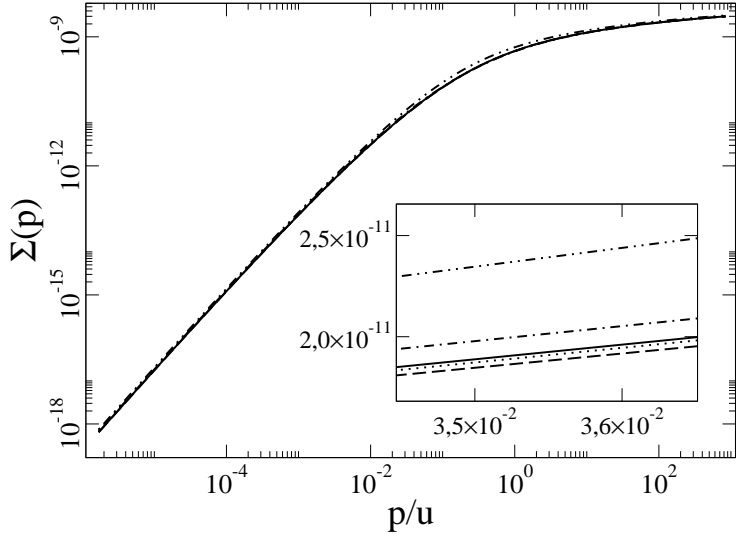}
\end{center}
\caption{ \label{comp_self_one} Comparison of the self-energy when expanding only 
the flow equations for the self-energy $\Sigma_\kappa(p;\tilde \rho)$ and its derivative $Z_\kappa(\tilde \rho)$ (strategy I): truncation is made at first (double dotted-dashed), second (dotted-dashed), third (dashed) and fourth (dotted) order; the complete solution is given by the straight line.
In the figure, $u=5.92 10^{-4} \Lambda$.}
\end{figure}

We have now all the ingredients to discuss the main goal of this paper: the analysis of the expansion 
of the three flow equations for $\mu_\kappa(\tilde \rho)$, 
$Z_\kappa(\tilde \rho)$ and $\Sigma_\kappa(p;\tilde \rho)$, eqs.~(\ref{eqmasa}-\ref{eqsigma}), 
around $\tilde \rho=0$. In doing so, one can write:
\beq
\Sigma_\kappa(p,\tilde \rho) = \sum_{n=0}^\infty \frac{1}{n!}\; \Sigma_\kappa^{(n)}(p) \; {\tilde \rho}^n.
\eeq
and
\beq
\chi_\kappa(\tilde \rho) = \sum_{n=0}^\infty \frac{1}{n!}\; \chi_\kappa^{(n)} \; {\tilde \rho}^n.
\eeq
together with eq.~(\ref{expmu}).
For example, when going to the first order, the six equations that have to be solved are:
\beq
\kappa\partial_\kappa\Sigma_\kappa^{(0)}(p)=-\frac{(1-\eta_\kappa/5)\Sigma_\kappa^{(1)}(p)}
{(1+\mu_\kappa^{(0)})^2}
\label{expsigma0}
\eeq
and
\begin{eqnarray}
\kappa\partial_\kappa\Sigma_\kappa^{(1)}(p)=(1+\eta_\kappa)\Sigma_\kappa^{(1)}(p)+
\frac{{2(\mu_\kappa^{(1)})}^2 Z_{\kappa}\kappa^2}{(1+\mu_\kappa^{(0)})^2}\left(f_\kappa(\tilde{p},0)
-\frac{2(1-\eta_\kappa/5)}{(1+\mu_\kappa^{(0)})}\right) \nonumber \\
+\frac{2f_\kappa(\tilde{p},0)}{(1+\mu_\kappa^{(0)})^2}\left(2\mu_\kappa^{(1)}\Sigma_\kappa^{(1)}(p)
+\frac{\Sigma_\kappa^{(1)}(p)^2}{\kappa^2Z_{\kappa}}\right)
+\frac{2(1-\eta_\kappa/5)\mu_\kappa^{(0)}\Sigma_\kappa^{(1)}(p)}{(1+\mu_\kappa^{(0)})^3},
\label{expsigma1}
\end{eqnarray}
which correspond to the expansion of eq.~(\ref{eqsigma}),
\beq
\kappa\partial_\kappa\chi_\kappa^{(0)}=\eta_\kappa\chi_\kappa^{(0)}-\frac{(1-\eta_\kappa/5)
\chi_\kappa^{(1)}}{(1+\mu_\kappa^{(0)})^2}
\label{expchi0}
\eeq
and
\beq
\kappa\partial_\kappa\chi_\kappa^{(1)}=(1+2\eta_\kappa)\chi_\kappa^{(1)}-\frac{2{(\mu_\kappa^{(1)})}^2}
{(1+\mu_\kappa^{(0)})^4}+\frac{10\mu_\kappa^{(0)}\chi_\kappa^{(1)}(1-\eta_\kappa/5)}{(1+\mu_\kappa^{(0)})^3},
\label{expchi1}
\eeq
which correspond to the expansion of eq.~(\ref{eqchi}), together with eqs.~(\ref{expmu0}) and (\ref{expmu1}).

In fact, it is possible to perform two kinds of expansion. First, in
order to isolate the effect of the field expansion just in the flow equations provided by the scheme presented 
in \cite{BMW}, we shall expand only the flow 
equations for  $\Sigma_\kappa(p;\tilde \rho)$ and its derivative at zero momenta
$Z_\kappa(\tilde \rho)$, eqs.~(\ref{eqsigma}) and (\ref{eqchi}), 
solving exactly the differential flow equation for 
$\mu_\kappa(\tilde \rho)$, eq.~(\ref{eqmasa}). 
For example, at first order, one should solve simultaneously eqs.~(\ref{expsigma0}-\ref{expsigma1}), 
(\ref{expchi0}-\ref{expchi1}), and (\ref{eqmasa}). This is called ``strategy I''. 
Second, to consider all the effects, we shall make the expansion in 
the three flow equations. For example, at first order, one should solve simultaneously eqs.~(\ref{expsigma0}-\ref{expsigma1}), 
(\ref{expchi0}-\ref{expchi1}), and (\ref{expmu0}-\ref{expmu1}). We call this ``strategy II''.
Notice that, as explained in \cite{BMW-num}, in order to get the correct scaling behavior it 
is mandatory to treat the equations for $Z_\kappa(\tilde \rho)$ and $\Sigma_\kappa(p;\tilde \rho)$ 
with the same 
approximations; it is then not possible to solve one of them completely while expanding the other one.

\begin{figure}[t]
\begin{center}
\includegraphics*[scale=0.4,angle=-90]{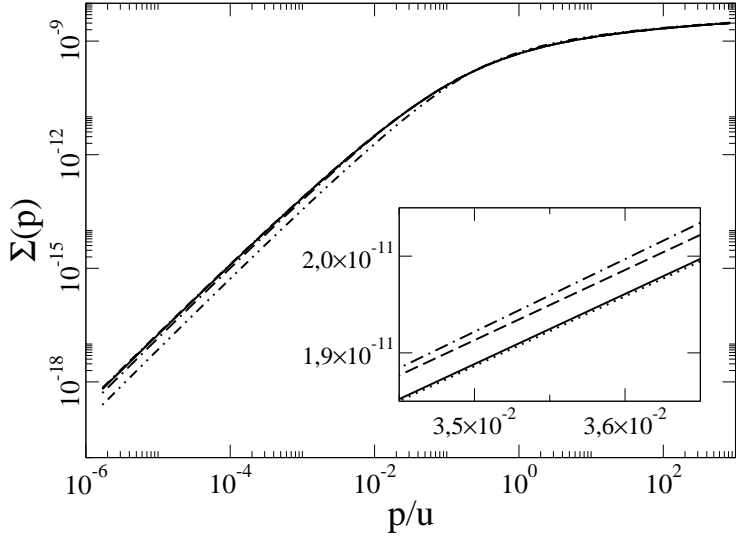}
\end{center}
\caption{ \label{comp_self_all} Comparison of the self-energy when expanding the three flow equations 
(strategy II): truncation 
is made at first (double dotted-dashed), second (dotted-dashed), third (dashed) and fourth (dotted) order; the 
complete solution is given by the straight line. In the figure, $u=5.92 10^{-4} \Lambda$.}
\end{figure}

Figure \ref{comp_self_one} presents the self-energy one gets truncating up to fourth order,
following strategy I; it is also shown the function obtained in \cite{BMW-num} 
(from now on, the latter function, obtained by solving 
 the 3 differential equations, eqs.~(\ref{eqmasa}), (\ref{eqchi}) and (\ref{eqsigma}), shall be called the ``complete solution'').
Figure \ref{comp_self_all} presents the same results when following strategy II. These Figures show that,
in both strategies of expansion, by truncating at first order one already gets a function with the correct 
shape in all momenta regimes.

\begin{figure}[t]
\begin{center}
\includegraphics*[scale=0.4]
{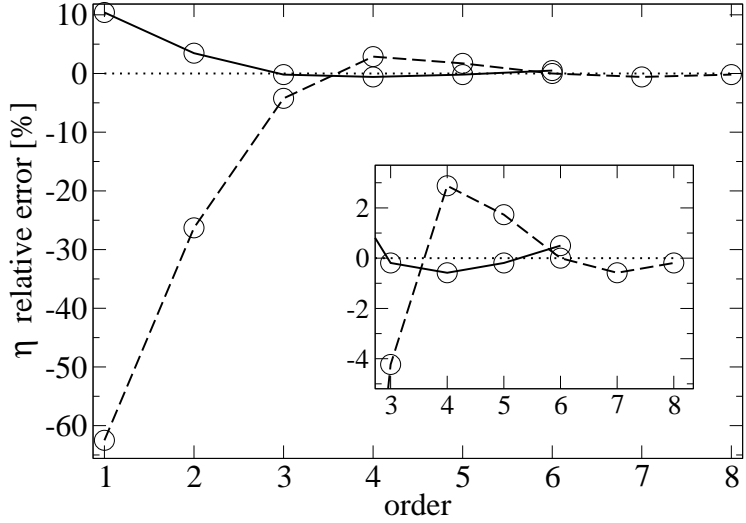}
\end{center}
\caption{ \label{eta} Relative error (measured in percent) for the anomalous dimension, with respect to 
the value coming from the complete solution, as a 
function of the truncation order. Full line: expanding  only the flow equations for the 
self-energy $\Sigma_\kappa(p;\rho)$ and its derivative $Z_\kappa(\rho)$ (strategy I); dashed line: expanding 
all flow equations (strategy II).}
\end{figure}

In order to make a quantitative evaluation of the approximate solution obtained doing 
the expansion, we have 
calculated different numbers describing the physical properties of the self-energy. 
First, as can be seen in both 
figures above, all solutions have, in the infrared ($p \ll u$), the potential behavior 
characterizing the scaling regime: 
$\Sigma(p) + p^2 \sim p^{2-\eta}$, where $\eta$ is the anomalous dimension. We have checked
that, at each order and in both strategies, the resulting self-energy does have scaling, and we 
extracted the corresponding value of $\eta$. In fact, this can be done in two different ways: either using the
$\kappa$--dependence of $Z_\kappa$ ($\eta=- \lim_{\kappa \to 0}\kappa \partial_\kappa \log Z_\kappa$) 
or the $p$--dependence
of $\Sigma(p)$ stated above. We checked that those two values always coincide, within numerical 
uncertainties. Figure \ref{eta} 
presents the relative error for $\eta$, at each order, when compared with the value following from the 
complete solution.  One observes: 1) in both strategies of expansion there is an apparent convergence, 
which is oscillatory; 2) the solution from strategy I reaches faster the correct result; 
3) when following strategy I, already with a second order truncation the error is about 
3\% and it drops to less that 1\% at the third order.
Nevertheless, due to the 
mixed characteristic of strategy I, when using this strategy at high order numerical problems arise:
indeed, this task demands the numerical evaluation of high order derivatives
of $\mu_\kappa(\tilde \rho)$, to be used in the various flow equations obtained when expanding that of  
$\Sigma_\kappa(p;\tilde \rho)$. If high precision in the result is required, strategy II is then 
numerically preferable.

It is important to observe here that the procedure which can be compared
to the usual truncation including a finite number of vertices is strategy II. Moreover, the inclusion of high order
vertices without performing any other approximation is difficult; for example, the
complete inclusion of the 6-point vertex has never been done. 
Accordingly, as can be seen in fig. \ref{eta}, when including only up to the 4-point vertex, as it usually done, the error in $\eta$ can be as large as 60\%.

\begin{figure}[t]
\begin{center}
\includegraphics*[scale=0.4]
{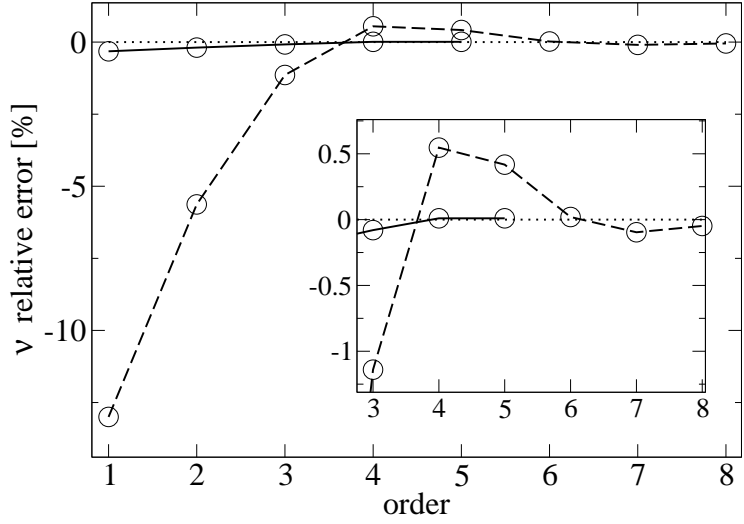}
\end{center}
\caption{ \label{nu} Relative error (measured in percent) for the critical exponent $\nu$, with respect to 
the value coming from the complete solution, as a 
function of the truncation order. Full line: expanding  only the flow equations for the 
self-energy $\Sigma_\kappa(p;\rho)$ and its derivative $Z_\kappa(\rho)$ (strategy I); dashed line: expanding 
all flow equations (strategy II).}
\end{figure}

A second number to assess the quality of the approximate solution is the critical exponent $\nu$.
In order to calculate it, we extract the renormalized dimensionful mass
from $m_R^2=\kappa^2\mu_\kappa(\tilde \rho=0)$ and we relate it to the microscopic one by
\begin{equation}
m_R^2(\kappa=0)\propto (m_R^2(\kappa=\Lambda)-m_{R,crit}^2(\kappa=\Lambda))^{2\nu},
\end{equation}
where $m_{R,crit}$ is the critical renormalized mass. With the complete solution one gets $\nu=0.647$, to be compared with the best accepted 
value \cite{zinn}: $\nu=0.6304\pm 0.0013$. Figure \ref{nu} presents the relative error of the value of $\nu$ extracted from the expansion. Once again, 
one observes that the convergence is much faster when following strategy I, i.e., when considering the effect 
of the expansion only on the self-energy equation.

The large momenta regime ($p \gg u$) of the self-energy can be calculated using perturbation theory, 
yielding the well 
known logarithmic shape: $\Sigma(p) \sim A \log(p/B)$, where $A$ and $B$ are constants. For the complete solution
presented in \cite{BMW-num} one can prove analytically that $A=u^2/9\pi^4$, which is only 8\% away from the exact result $A=u^2/96\pi^2$ (please observe that this coefficient is given by a 2-loop diagram for the self energy which is only approximatively included at this order). The proof of this analytical result remains 
valid when performing the field expansion, at any order and within both 
strategies. We have checked that our numerical solution always has the correct shape, with $A=u^2/9\pi^4$.
This is due to the fact that already the first order in the expansion of $\Sigma_\kappa(p;\tilde\rho)$ 
around $\tilde \rho=0$ contains the same 2-loop diagrams contributing to the complete solution.

\begin{figure}[t]
\begin{center}
\includegraphics*[scale=0.4]
{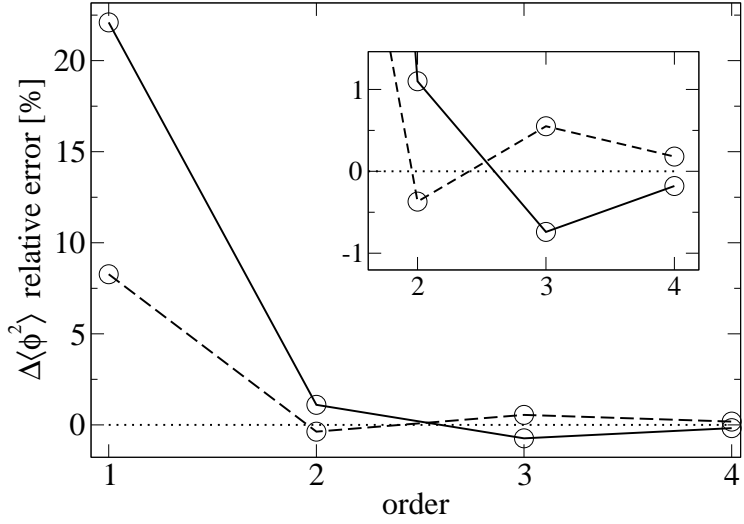}
\end{center}
\caption{\label{coefc} Relative error (measured in percent) for $\Delta\langle\phi^2\rangle$, with respect to 
the value coming from the complete solution, as a 
function of the truncation order. Full line: expanding  only the flow equations for the 
self-energy $\Sigma_\kappa(p;\rho)$ and its derivative $Z_\kappa(\rho)$ (strategy I); dashed line: expanding 
all flow equations (strategy II).}
\end{figure}

In order to study the quality of the self-energy in the intermediate momenta regime, we have calculated a 
quantity which is very sensitive to this cross-over region:
\beq\label{integralc} 
\Delta\langle
\phi^2\rangle= 
\int\frac{d^3 
p}{(2\pi)^3}\,\left(
\frac{1}{p^2+\Sigma(p)}-\frac{1}{p^2}\right).
\eeq
(the integrand is non zero only in the region $10^{-3} \simle p/u \simle 
10$, see for example \cite{Blaizot:2004qa}). This quantity received 
recently much attention because
it has been shown \cite{JPTc} that for a scalar model with $O(N)$ symmetry,
in $d=3$ and  $N=2$, it determines the shift of the critical
temperature of the weakly repulsive Bose gas. It has then been
widely evaluated by many methods, for different values of $N$, in
particular, for  $N=1$. With the numerical solution found in \cite{BMW-num}, one gets
a number almost within the error bars of the best
accepted results available in the literature, using lattice and 
7 loops resumed perturbative calculations. Please observe that these errors are as large as 10\%, which
is an indication that this quantity is particularly difficult to calculate.
In Figure \ref{coefc} we plot the relative error
in $\Delta\langle\phi^2\rangle$, at each order of the expansion, when compared with the complete solution 
result found in \cite{BMW-num}. One can appreciate that 
1) for both expansion strategies there is an apparent convergence, which is also oscillatory; 
2) in both strategies, already with a second order truncation the error is about 1\%. 

\section{Summary and Conclusions}

In this article, the inclusion of a finite number of vertices in NPRG flow equations is analyzed.
An unsolved difficulty of this usual strategy (originally proposed by Weinberg) is the estimation of the error introduced at a given step. Moreover, without performing further approximations, it is very hard to reach high orders of the procedure. The study of its convergence is thus a difficult task. 
In the present work we analyse this problem using a different approximation scheme \cite{BMW}: instead of considering a finite number of vertices, this procedure includes all of them, although approximately. Within this context, it is possible to estimate the error of the Weinberg approximation, order by order.
To do so one can perform, on top of the approximation presented in \cite{BMW}, the usual truncation in the number of vertices. The analysis has been done in the particular case of the 2-point function of the scalar field theory in $d=3$ at criticality. It has been shown \cite{BMW-num} that, at least in this case, the procedure proposed in \cite{BMW} yields very precise results. 
Another interesting outcome of the present work follows from the fact that, within the approximation \cite{BMW}, truncation in the number of vertices is equivalent to an expansion in powers of a constant external field. The latter is usually employed in the DE context in order to deal with complicated situations. The analysis of the present paper generalizes this expansion procedure when non zero external momentum are involved.

The calculation of the $2$-point function demands the study of both the $p=0$ and the $p\neq 0$ sectors. 
While the first one is given by the well studied DE flow equations, the latter 
follows from the approximation scheme introduced in \cite{BMW} to calculate the flow of $\Sigma_\kappa(p;\rho)$.
We used two different strategies to perform the field expansion, both of them around zero external field: 
either expanding only the flow equation for the self-energy (and its derivative) (strategy I), or
both the effective potential and the self-energy (and its derivative) flow equations (strategy II). 
We have studied the convergence of various quantities measuring physical 
properties of the self-energy in all momenta regimes: the critical exponents $\eta$ and $\nu$ of the 
infrared regime, 
the coefficient of the ultraviolet logarithm, and $\Delta\langle\phi^2\rangle$ which is dominated by the 
crossover momenta regime. 

As stated in section \ref{expansion}, the strategy that can be compared to the usual truncation which includes a finite number of vertices is strategy II. For example, including completely the $4$-point vertex as it is usually done (i.e., in the language of field expansion, going only up to the first order of the expansion), when describing the deep infrared regime one could make errors as big as
60\% in the critical exponent $\eta$ (see figure \ref{eta}). If one wants results with less that 5\% error for this quantity, the inclusion of up to $8$-point vertices (i.e., going up to third order) is necessary. 

However, when going to higher orders in the field expansion, the series for all considered quantities seem to converge rapidly, within both strategies.
The convergence is faster when using strategy I, i.e., when making the expansion only for the approximate flow equation
resulting from the method presented in \cite{BMW}. For example, using strategy I, a third order truncation introduces a relative error smaller that 1\% for all studied quantities;
while using strategy II, in order to reach the same error one needs 6th order for $\eta$, 4th order for $\nu$ and 2nd order for $\Delta\langle\phi^2\rangle$.
Nevertheless, due to numerical difficulties, if trying to go to high order expansions, it is preferable to use strategy II, i.e., expanding also the effective potential flow equation.

It is difficult to assess the generality of these results on the use of field expansion on top of the strategy proposed in \cite{BMW}. Of course, there are situations where expanding in an external field is not a priori convenient. One can mention as a first example, situations where there is a physical external field (as in a broken phase or when an external source for the field is considered). A second example is two-dimensional systems where even in the DE, the field expansion does not seem to converge. Nevertheless, the short study presented in the present paper allows to consider field expansion on top of the approximation proposed in \cite{BMW} as a possible strategy to deal with many involved models, as for example QCD.

\bibliographystyle{unsrt}

\end{document}